\documentclass{article}%
\usepackage{amsmath}
\usepackage{amsfonts}
\usepackage{amssymb}
\usepackage{graphicx}%
\makeatletter
\providecommand{\LyX}{L\kern-.1667em\lower.25em\hbox{Y}\kern-.125emX\@}
\makeatother
\begin{document}

\title{Maximally-Disordered Distillable Quantum States }
\author{Somshubhro Bandyopadhyay\thanks{Present address: Department of Chemistry,
University of Toronto, 80 St. George St., Toronto, ON, M5S3H6, Canada
Email:sbandyop@chem.utoronto.ca} \ and Vwani Roychowdhury
\thanks{Email:vwani@ee.ucla.edu}\\{\normalsize Department of Electrical Engineering, UCLA, Los Angeles, CA
90095}}
\date{}
\maketitle

\begin{abstract}
{\small We explore classical to quantum transition of correlations by studying
the quantum states located just outside of the
classically-correlated-states-only neighborhood of the maximally mixed state
(the largest separable ball (LSB)). We show that a natural candidate for such
states raises the possibility of a layered transition, i.e., an annular region
comprising only classical and the classical-like bound entangled states,
followed by free or distillable entanglement. Surprisingly, we find the
transition to be abrupt for bipartite systems: distillable states emerge
arbitrarily close to the LSB. For multipartite systems, while the radius of
the LSB remains unknown, we determine the radius of the largest undistillable
ball. Our results also provide an upper bound on how noisy shared entangled
states can be for executing quantum information processing protocols. }

\end{abstract}



\noindent\newline\textbf{Introduction}: The states of a bipartite quantum
system can either be separable or inseparable: In the former, the correlations
between the subsystems are entirely classical (i.e., the correlations can be
mimicked by a local description of the individual subsystems), and the states
can be written as a convex combination of product states:
\begin{equation}
\rho^{AB}=\sum p_{i}\rho_{i}^{A}\otimes\rho_{i}^{B}, \label{Eq1}%
\end{equation}
where $A,B$ denote the subsystems and the positive probabilities $p_{i\text{
\ }}$ add up to one. The density matrices $\rho_{i}$ act on the individual
Hilbert spaces corresponding to the subsystems $A$ and $B.$ If a state cannot
be written in the form (1) it is said to be inseparable or entangled and can
exhibit correlations that cannot be explained by any classical theory
\cite{entanglement}. Recent advances have led to discoveries of two
qualitatively different forms of entanglement, namely, bound entanglement
\cite{pawel,peresbruss,horo3,ibm2} and distillable entanglement
\cite{popescu,deutsch,ibm1,gisin}

It is of fundamental importance how the different classes of quantum states
are distributed in the Hilbert space. In particular, an understanding of the
distribution of different states along the boundaries in the Hilbert space,
where transitions from classically correlated states to quantum correlated
states occur, has the potential to provide new insights into the very nature
of quantum non-locality. A legitimate starting point would be to obtain a
picture of such a distribution relative to the maximally mixed state: the
state which is completely disordered (both locally and globally) and does not
have any correlation between the subsystems. It is natural to expect that as
one starts to move away from the maximally mixed state one would encounter
more ordered states, as well as emergence of correlations between the
subsystems. It has been shown that the neighbourhood of a maximally mixed
state consists entirely of separable states \cite{sanpera,vidal,braunstein},
and recently the exact size of this separable neighbourhood for bipartite
systems has been obtained \cite{gurvits}. Thus, transitions from classical
correlations to quantum can be observed just beyond the boundary of the LSB.

While determining the exact size of the separable neighbourhood constitutes a
major progress, answers to several questions related to the characterization
and distribution of the inseparable states beyond the LSB remain unknown,
primarily because of the existence of two very different types of
entanglement, viz., bound entanglement (BE) and distillable entanglement. The
former class, though quantum correlated, is qualitatively very similar to
classically correlated states because of their positivity under partial
transposition (PPT) \cite{peres, horo1} and usefulness that is no better than
classically correlated states for reliable quantum information processing. The
latter class, however, is negative under partial transposition (NPT) and shows
distinct and demonstrable non-local features in quantum communication
protocols \cite{teleportation,cryptography,sdc} \ \footnote{We also note that
the existence of BE state that are NPT has been conjectured by several groups,
but a rigorous proof is still pending.}.

We first ask how one may construct entangled states that are closest to the
LSB. It was often thought that the entangled states closest to the maximally
mixed state can be reached via mixing of an entangled state with the maximally
mixed state, e.g., the pseudo-pure states used in room temperature NMR quantum
computing. We will refer to this class of states as the Werner-type states
\cite{werner}. We show that while this is true for the simplest case of
$2\otimes2$, for all other bipartite systems, \emph{there is a finite gap
between the boundary of the LSB and the nearest inseparable state (which turns
out to be distillable as well)} that can be reached via such mixing.

Such a finite gap between the most-disordered distillable Werner-type states
and the LSB opens up several interesting possibilities. For instance, we can
think of a PPT neighbourhood of the maximally mixed state distinct from the
LSB within which there is no NPT state. Moreover, there can as well be a
largest undistillable neighbourhood distinct from both the largest separable
and the largest PPT ones, with bound entangled states as the only inseparable
states in the intermediate regime. We therefore ask \emph{at what distances
and in what order, if any, relative to the boundary of the LSB can distillable
and bound entangled states be found}? 

For bipartite quantum systems, we show, using explicit constructions, that
instead of the different types of inseparable states appearing in a layered
fashion, both distillable and NPT $n$-copy undistillable states (conjectured
to be NPT bound entangled) \cite{bandyoroychowdhury,ibm3} can be found
arbitrarily close to the boundary of the LSB. \emph{The class of distillable
states that lie arbitrarily close to the boundary of the LSB} are therefore,
\emph{the maximally-disordered distillable states}. Such states are
constructed via perturbations of the separable states on the surface of the
LSB. A subset of such separable states on the surface are \textit{the
maximally mixed states of }$(D-1)$ \textit{dimensional subspaces }where $D$ is
the dimension of the full Hilbert space..

We also provide constructions for maximally disordered distillable states in
multipartite quantum systems, where the exact size of the LSB is still
unknown. However, using a result obtained in Ref. \cite{gurvits} we are able
to show that all quantum states that are on or inside the ball of radius
$R_{LSB}$ ($D$ is now the \emph{total dimension} of the multipartite system)
are PPT and therefore represents the largest undistillable ball in the
multipartite case. Proceeding in a similar way as in the bipartite case we
construct multipartite distillable states that are arbitrarily close to this
PPT ball. Interestingly this shows that \textit{given a composite quantum
system, the radius of the largest undistillable neighbourhood is independent
of all possible partitions and depends only on the total dimension}.
Remarkably, the preceding result coupled with the conjecture that the the LSB
for multipartite systems is smaller than that for bipartite systems, lead to a
\emph{new conjecture that the transition for general multipartite systems is
indeed layered.} \newline\textbf{Minimum distance of the Werner-type entangled
states from the maximally mixed state:} In this paper, the distance between
any two density matrices, $\rho$ and $\sigma,$ will be given by the
Hilbert-Schmidt distance defined as: $\delta=\left\Vert \rho-\sigma\right\Vert
\equiv\sqrt{tr\left(  \left(  \rho-\sigma\right)  ^{2}\right)  }.$ We begin by
considering the class of bipartite mixed states , $\rho_{x}=x\rho+\frac
{(1-x)}{D}\mathbf{I}$, where $x>0$ and $\rho$ is any density matrix in
$d\otimes d$ . The total dimension of the composite Hilbert space is denoted
by $D=d^{2}.$ Obviously if $\rho_{x}$ is entangled for some values of $x,$
then $\rho$ must be an entangled state. Entanglement properties of such states
are of considerable importance because the generic quantum states used in NMR
quantum computing, the so called pseudo-pure states \cite{nmr}, and the famous
Werner states \cite{werner} are of similar form. In what follows we answer the
question, are the states $\rho_{x},$ when inseparable, the closest inseparable
states to the maximally mixed state? In Ref \cite{gurvits} it was shown that
the states $\rho_{x}$ are always separable when $x\leq\frac{1}{D-1}$.
Moreover, $\rho_{x}$ becomes inseparable when $\rho$ is taken to be a
maximally entangled state of Schmidt rank two and $x>\frac{2}{D+2}$. The proof
of inseparability follows from the fact that when $\rho$ is chosen to be a
maximally entangled state of Schmidt rank 2, then $\rho_{x}$ becomes an NPT
state for $x>\frac{2}{D+2} $ \cite{gurvits}.

We first show that \emph{$\rho_{x}$ indeed becomes distillable} when
$x>\frac{2}{D+2}$, and $\rho$ is one of the four Bell states, $\left\{
\left\vert \Phi^{\pm}\right\rangle ,\left\vert \Psi^{\pm}\right\rangle
\right\}  $ defined in the standard basis as $\left\vert \Phi^{\pm
}\right\rangle =\frac{1}{\sqrt{2}}\left(  \left\vert 00\right\rangle
\pm\left\vert 11\right\rangle \right)  $ and $\left\vert \Psi^{\pm
}\right\rangle =\frac{1}{\sqrt{2}}\left(  \left\vert 01\right\rangle
\pm\left\vert 10\right\rangle \right)  $. Let $\rho=\left\vert \Phi
^{+}\right\rangle \left\langle \Phi^{+}\right\vert $. Then $\left\langle
\Psi^{-}\right\vert (x\rho^{PT}+\frac{(1-x)}{D}\mathbf{I})\left\vert \Psi
^{-}\right\rangle <0$ when $x>\frac{2}{D+2}.$ Since $\left\vert \Psi
^{-}\right\rangle $ is a Schmidt rank two state, the state $\rho_{x}$ is
therefore distillable\footnote{A bipartite quantum state $\rho$ is said to be
distillable iff there exists an integer $n$ and a Schmidt rank two state
$\left\vert \psi\right\rangle $ such that $\left\langle \psi\right\vert
\left(  \rho^{PT}\right)  ^{\otimes n}\left\vert \psi\right\rangle <0$
\cite{horo3}.}.

Let us now note that when $x\leq\frac{1}{D-1},$ the distance of the state
$\rho_{x}$ from the maximally mixed state is always less than or equal to
$R_{LSB}$ and the equality is achieved when $\rho$ is any pure state and
$x=\frac{1}{D-1}.$ The intermediate regime $\frac{1}{D-1}<x\leq\frac{2}{D+2},$
therefore corresponds to the states that are outside the LSB and the
inseparability/separability property of these states remains to be determined.

We now show that \emph{if $x\leq\frac{2}{D+2}$, then states of the form $
\rho_{x}=x\rho+\frac{(1-x)}{D}\mathbf{I}$ are always separable for all $\rho$.
} It was proved in Ref \cite{vidal}\ that a full rank mixed state is separable
if the minimum eigenvalue is greater or equal to $\frac{1}{D+2}.$ We show that
if $x\leq\frac{2}{D+2}$ then for any $\rho$ the minimum eigenvalue of
$\rho_{x}$ is always greater or equal to $\frac{1}{D+2}$ from which our
conclusion immediately follows. Let us first suppose that $\rho$ is not a full
rank state. Then the minimum eigenvalue of $\rho_{x}$ is $\frac{(1-x)}{D}%
\geq\frac{1}{D+2}$ if $x\leq\frac{2}{D+2}$. Now assume that $\rho$ is a full
rank state and the minimum eigenvalue of $\rho$ is $\lambda.$ Then the minimum
eigenvalue of $\rho_{x}$ is $\lambda x+\frac{(1-x)}{D}$. Then $\lambda
x+\frac{(1-x)}{D}\geq\frac{1}{D+2}$, which implies that $x\leq\frac
{2}{(D+2)(1-D\lambda)}$. Since $D\lambda\leq1,$ we have $\frac{2}%
{(D+2)(1-D\lambda)}>\frac{2}{D+2}.$

Now note that the distance of the state $\rho_{x}$ from the maximally mixed
state is given by $x\sqrt{Tr\left(  \rho^{2}\right)  -\frac{1}{D}}.$ Thus, the
inseparable states (as shown earlier, such states are also distillable)
nearest to the maximally mixed state that can be reached via perturbation can
be expressed as a mixture of a pure entangled state and the maximally mixed
state and therefore at a distance $R=\frac{2}{D+2}\sqrt{\frac{(D-1)}{D}}$. We
note that $R\geq R_{LSB}$ \ and the equality is achieved only when $D=4,$
corresponding to the systems in $2\otimes2.$ From the ratio $\frac{R}{R_{LSB}%
}=2\left(  \frac{D-1}{D+2}\right)  $ implies that as the dimension of the
Hilbert space increases, the entangled states that can be reached via
perturbation start to move away farther from the boundary of the LSB, and the
ratio becomes as large as $2$. \newline\textbf{Constructions of distillable states arbitrarily close to
the boundary of the LSB:} We first note that the radius of the largest
separable ball is, in fact, the distance between the maximally mixed state and
the maximally mixed state in any $(D-1)$ dimensional subspace, i.e.,
$R_{LSB}=\left\Vert \frac{1}{D}\mathbf{I}-\frac{1}{D-1}\mathbf{I}%
_{D-1}\right\Vert $, where $\mathbf{I}_{D-1}=\frac{1}{D-1}\left(
\mathbf{I}-\left\vert \varphi\right\rangle \left\langle \varphi\right\vert
\right)  $, for some pure state $\left\vert \varphi\right\rangle $. This turns
out to be a useful observation. We now focus our attention on a class of
operators that are the partial transposition of $\frac{1}{D-1}\mathbf{I}%
_{D-1}$:
\begin{equation}
\sigma_{\varphi}(k)=\frac{1}{D-1}\left(  \mathbf{I}-\left\vert \varphi
\right\rangle \left\langle \varphi\right\vert \right)  ^{PT}, \label{Eq2}%
\end{equation}
where $\left\vert \varphi\right\rangle $ is a pure state of Schmidt rank $k$,
$k=1,\cdots,d$, and is of the form $\sum_{i=0}^{i=k-1}\eta_{i}\left\vert
ii\right\rangle $ where $\eta_{i}^{\prime}s$ are real and positive and
$\sum_{i=0}^{k-1}\left\vert \eta_{i}\right\vert ^{2}=1$. The superscript $PT$
denotes partial transposition. One can now check that the operator
$\sigma_{\varphi}(k)$ is indeed a density matrix. Firstly, it has trace one,
since the trace of it's partial transpose is one and trace is invariant under
partial transposition. Now note that the eigendecomposition of the operator
$\left\vert \varphi\right\rangle ^{PT}\left\langle \varphi\right\vert $ is
given by
\begin{align}
\left(  \left\vert \varphi\right\rangle \left\langle \varphi\right\vert
\right)  ^{PT}  &  =\textstyle\sum\limits_{i=0}^{k-1}\left\vert \eta
_{i}\right\vert ^{2}\left\vert ii\right\rangle \left\langle ii\right\vert
+\textstyle\sum\limits_{i,j=0,i<j}^{k-1}\eta_{i}\eta_{j}\left\vert \psi
_{ij}^{+}\right\rangle \left\langle \psi_{ij}^{+}\right\vert \nonumber\\
&  \hbox{\hspace*{1cm}}-\textstyle\sum\limits_{i,j=0,i<j}^{k-1}\eta_{i}%
\eta_{j}\left\vert \psi_{ij}^{-}\right\rangle \left\langle \psi_{ij}%
^{-}\right\vert , \label{Eq3}%
\end{align}
where $\left\vert \psi_{ij}^{\pm}\right\rangle =\frac{1}{\sqrt{2}}\left(
\left\vert ij\right\rangle \pm\left\vert ji\right\rangle \right)  .$Note that
the eigenvectors of $\left(  \left\vert \varphi\right\rangle \left\langle
\varphi\right\vert \right)  ^{PT}$ span $k^{2}$ dimensional subspace. Now
substituting (\ref{Eq3}) in (\ref{Eq2}) one can see that the operator is
Hermitian and positive semidefinite. Therefore it is a density matrix.

When $k=1$, $\left\vert \varphi\right\rangle $ is a product state. Hence,
$\sigma_{\varphi}(1)$ is the maximally mixed state of a $(D-1)$ dimensional
subspace and is on the surface of the LSB. The same cannot be said when
$k\geq2$ because $\left\vert \varphi\right\rangle $ is now a pure entangled
state. In fact we next show that the states, $\sigma_{\varphi}(k)$ are of full
rank and also lie on the surface of the LSB for all $k\geq2$. They are
therefore separable.

Substituting (\ref{Eq3}) in (\ref{Eq2}) one can obtain the spectral
decomposition of $\sigma_{\varphi}(k)$,
\begin{align}
\sigma_{\varphi}(k)  &  = \frac{1}{D-1}\left(  \textstyle\sum\limits_{i=0}%
^{k-1}\left(  1-\left\vert \eta_{i}\right\vert ^{2}\right)  \left\vert
ii\right\rangle \left\langle ii\right\vert \right. \nonumber\\
&  \hspace*{-1cm} + \textstyle\sum\limits_{i,j=0,i<j}^{k-1}\left(  1-\eta
_{i}\eta_{j}\right)  \left\vert \psi_{ij}^{+}\right\rangle \left\langle
\psi_{ij}^{+}\right\vert \nonumber\\
&  \hspace*{-1cm} \left.  + \textstyle\sum\limits_{i,j=0,i<j}^{k-1}\left(
1+\eta_{i}\eta_{j}\right)  \left\vert \psi_{ij}^{-}\right\rangle \left\langle
\psi_{ij}^{-}\right\vert +\mathbf{I}_{D-k^{2}} \right)  , \label{Eq4}%
\end{align}
where $\mathbf{I}_{D-k^{2}}$ is the projector on the $(D-k^{2})$ dimensional
subspace. One can now see that $\sigma_{\varphi}(k)$ is of full rank, because
none of the eigenvalues is zero. A calculation of the distance from the
maximally mixed state leads to the result that the distance $\left\Vert
\frac{1}{D}\mathbf{I}-\sigma_{\varphi}(k)\right\Vert =R_{LSB}$ for all
$k=2,\cdots,d$. Note that \emph{the distance is independent of $k$}, and the
states are separable because they lie on the surface of the LSB.

We now consider the class of states obtained via perturbation of the states $
\sigma_{\varphi}(k)$,
\begin{equation}
\rho_{\psi,\varphi}\left(  \epsilon,k\right)  =\epsilon\left\vert
\psi\right\rangle \left\langle \psi\right\vert +\left(  1-\epsilon\right)
\sigma_{\varphi}\left(  k\right)  , \label{Eq5}%
\end{equation}
where $\left\vert \psi\right\rangle $ is a pure entangled state to be fixed
later and $0<\epsilon<1$. We next prove that \emph{for the cases, k=1,2}
\ \emph{there exist states, $\left\vert \psi\right\rangle $, such that
$\rho_{\psi,\varphi}\left(  \epsilon,k\right)  $ are distillable} for some
finite range of $\epsilon.$In the rest of this discussion, we will ignore the
subscript $\varphi$ in the cases when it is apparent from the context.

\textbf{k=1:} Without loss of generality, we can write $\rho_{\psi}\left(
\epsilon,1\right)  =\epsilon\left\vert \psi\right\rangle \left\langle
\psi\right\vert +\frac{1-\epsilon}{D-1}\mathbf{I}_{D-1}$, where $\frac
{1-\epsilon}{D-1}\mathbf{I}_{D-1}=\frac{1-\epsilon}{D-1}(\mathbf{I}-\left\vert
00\right\rangle \left\langle 00\right\vert )$. Let $\left\vert \psi
\right\rangle $ be a pure entangled state of Schmidt rank two: $\left\vert
\psi\right\rangle =\frac{1}{\sqrt{6}}\left(  \left\vert 10\right\rangle
-\left\vert 11\right\rangle -2\left\vert 01\right\rangle \right)  .$To prove
that $\rho_{\psi}\left(  \epsilon,1\right)  $ is distillable, we need to show
that there exists a Schmidt rank two state $\left\vert \chi\right\rangle $
such that $\left\langle \chi\right\vert \rho^{PT}\left\vert \chi\right\rangle
<0$. Construct the following state
\begin{equation}
\left\vert \chi\right\rangle =\alpha\left\vert 00\right\rangle +\frac{\beta
}{2}\left(  \left\vert 10\right\rangle -\left\vert 11\right\rangle -\left\vert
00\right\rangle +\left\vert 01\right\rangle \right)  , \label{chi}%
\end{equation}
where $\alpha=\left\vert \alpha\right\vert e^{i\vartheta_{1}}$ and
$\beta=\left\vert \beta\right\vert e^{i\vartheta_{2}}$ are complex quantities.
By showing that the reduced density matrices are of rank 2, one can easily
verify that the state $\left\vert \chi\right\rangle $ is indeed of Schmidt
rank two. We choose $\vartheta_{1},\vartheta_{2}$ such that $\vartheta
_{1}-\vartheta_{2}=\pi$. With this constraint, the normalization condition for
$\left\vert \chi\right\rangle $ reads as $\left\vert \alpha\right\vert
^{2}+\left\vert \beta\right\vert ^{2}+\left\vert \alpha\right\vert \left\vert
\beta\right\vert =1.$ One can now easily check that: $\left\langle
\chi\right\vert \rho_{\psi}^{PT}\left(  \epsilon,1\right)  \left\vert
\chi\right\rangle =-\frac{\epsilon}{3}\left\vert \alpha\right\vert \left\vert
\beta\right\vert +\frac{3(1-\epsilon)}{4(D-1)}\left\vert \beta\right\vert
^{2},$where we have used $\vartheta_{1}-\vartheta_{2}=\pi$. \ Let
$\frac{\left\vert \alpha\right\vert }{\left\vert \beta\right\vert }=y.$
Requiring $\left\langle \chi\right\vert \rho_{\psi}^{PT}\left(  \epsilon
,1\right)  \left\vert \chi\right\rangle <0$ for distillability, one obtains
\begin{equation}
\epsilon>\frac{1}{1+yA}, \label{cond}%
\end{equation}
where $A=\frac{4(D-1)}{9}$ is a constant. The above condition can be always
satisfied for every $\epsilon>0$ as we can appropriately choose the ratio,
$y$, and then solve for $\left\vert \alpha\right\vert $ and $\left\vert
\beta\right\vert $ by solving: $\left\vert \alpha\right\vert ^{2}+\left\vert
\beta\right\vert ^{2}+\left\vert \alpha\right\vert \left\vert \beta\right\vert
=1$. Hence, \emph{$\rho_{\psi,\varphi}$ is distillable}.

\textbf{k=2: }Let $\rho_{\psi}\left(  \epsilon,2\right)  =\epsilon\left\vert
\psi\right\rangle \left\langle \psi\right\vert +\left(  1-\epsilon\right)
\sigma_{\varphi}(2)$ where $\sigma_{\varphi}\left(  2\right)  =\frac{1}%
{D-1}\left(  \mathbf{I}-\left\vert \Phi^{+}\right\rangle \left\langle \Phi
^{+}\right\vert \right)  ^{PT}$. That is, we have chosen $\left\vert
\varphi\right\rangle $ to be the the Bell state $\left\vert \Phi
^{+}\right\rangle $. We now choose $\left\vert \psi\right\rangle $ as the
singlet state. Therefore, the state under consideration is: $\epsilon
\left\vert \Psi^{-}\right\rangle \left\langle \Psi^{-}\right\vert
+\frac{1-\epsilon}{D-1}\left(  \mathbf{I}-\left\vert \Phi^{+}\right\rangle
\left\langle \Phi^{+}\right\vert \right)  ^{PT}$. By noting that $\left\vert
\Phi^{+}\right\rangle $ is the eigenvector corresponding to the negative
eigenvalue of $\left(  \left\vert \Psi^{-}\right\rangle \left\langle \Psi
^{-}\right\vert \right)  ^{PT}$ it follows, $\left\langle \Phi^{+}\right\vert
\rho^{PT}\left(  \epsilon,2\right)  \left\vert \Phi^{+}\right\rangle
=-\frac{\epsilon}{2}$. Hence, the state $\rho_{\psi,\varphi}\left(
\epsilon,2\right)  $ is distillable when $\epsilon>0$.

The preceding results show that one can have distillable states arbitrarily
close to the surface of the LSB as the states $\sigma_{\varphi}(k)$ are on the
surface of the LSB. Recently, new classes of NPT states that are
\textit{\ $n$-copy undistillable} for any $n\geq1$ have been obtained in
Ref.\cite{bandyoroychowdhury} One such class corresponds to the states that
are of the form (\ref{Eq5}), where $k\geq3.$ Hence, we also have
\textit{$n$-copy undistillable} states \emph{on the immediate boundary} of the
LSB since $\sigma_{\varphi}\left(  k\right)  $ lies on the LSB for all
$k\geq1$. \newline\textbf{Constructions for multipartite systems:} Some of our
results and also that of Ref \cite{gurvits} can be directly applied to
multiparty systems but with some caution. A multipartite state can be
separable in all bipartite cuts, but can still be inseparable. Such true bound
entangled states do exist \cite{ibm2}. However, positivity/negativity of a
multipartite state under partial transposition under all possible bipartite
partitions do indeed provide useful information regarding distillability. For
instance, a multipartite state cannot be distillable, and at the same time,
PPT/separable across \emph{all} bipartite cuts. Moreover, if we can show that
a multipartite state has distillable entanglement across at least one
bipartite cut, then the state is distillable. We now show that \emph{the
boundary of the ball of radius $\frac{1}{\sqrt{D(D-1}}$} around the maximally
mixed state \emph{contains distillable multipartite states}, where $D$ is now
the total dimension of the multipartite system. The construction is very
similar to that in the case of bipartite systems. We provide it for the
$2\otimes2\otimes2$ case, but it can be trivially generalized for higher
dimensions. Consider the following class of states%
\begin{equation}
\widetilde{\rho}\left(  \epsilon,k\right)  _{123}=\epsilon\left\vert
\psi\right\rangle _{123}\left\langle \psi\right\vert +\left(  1-\epsilon
\right)  \widetilde{\sigma}_{123}, \label{mpartite}%
\end{equation}
where
\begin{equation}
\left\vert \psi\right\rangle _{123}=\frac{1}{\sqrt{6}}\left(  \left\vert
100\right\rangle -\left\vert 111\right\rangle -2\left\vert 011\right\rangle
\right)  _{123} \label{mpartitepsi}%
\end{equation}
and $\widetilde{\sigma}=\frac{1}{7}\left(  \mathbf{I}-\left\vert
000\right\rangle \left\langle 000\right\vert \right)  _{123}$. To show that
the state (\ref{mpartite}) is inseparable/distillable, it is sufficient to
show that the state is inseparable across at least one bipartite cut. Let us
consider the $1:(2,3)$ cut, and denote the states of the qubits 2 and 3 as
follows: $\left\vert 00\right\rangle _{23}=\left\vert \overline{0}%
\right\rangle _{23},\left\vert 11\right\rangle _{23}=\left\vert \overline
{1}\right\rangle _{23},\left\vert 01\right\rangle _{23}=\left\vert
\overline{2}\right\rangle _{23},\left\vert 10\right\rangle _{23}=\left\vert
\overline{3}\right\rangle _{23}$. Substituting these in the above two
equations, we get exactly the same distillable states that we have used in the
bipartite case.

We close our discussion of the multipartite case with the following claim:
\emph{For a multipartite quantum system of total dimension $D,$ the exact
radius of the largest PPT ball around the maximally mixed state is given by
$\frac{1}{\sqrt{D(D-1}}.$ It is also the largest undistillable ball.} First,
using the result of Ref \cite{gurvits}, we observe that for all possible
bipartite cuts, the largest separable ball has the radius $\frac{1}%
{\sqrt{D(D-1}}.$ Next, if a state is NPT, then it has to be NPT in one of the
bipartite cuts and therefore, it lies outside the ball of radius $\frac
{1}{\sqrt{D(D-1}}.$ Hence, the radius of the largest PPT ball is at least
$\frac{1}{\sqrt{D(D-1}}$. However, we have just shown that on the boundary of
such a ball there exists distillable states. Hence, the largest PPT ball and
the largest undistillable ball are the same for the multipartite case, with a
radius of $\frac{1}{\sqrt{D(D-1}}$.\newline\textbf{Discussions: }We have shown
that the inseparable states nearest to the maximally mixed state can never be
reached via mixing of an entangled state with the maximally mixed state other
than in $2\otimes2$. We have also addressed the issue of distribution of
different types of inseparable states on the boundary of the LSB for the
bipartite case and have shown via explicit constructions that distillable
states exist arbitrarily close to the surface of the LSB. Hence, these states
are the \emph{maximally-disordered }distillable states. Thus, for the
bipartite case, our results show that the LSB, the largest PPT ball, and the
largest undistillable balls all have the same radius. In the case of
multipartite systems, immediate extensions of the results for the bipartite
case imply that the radius of the largest PPT ball and the largest
undistillable ball are the same. This shows that for any composite quantum
system, the radius of the largest undistillable ball around the maximally
mixed state is independent of all possible partitions and depends only on the
total dimension. However, it is likely that there is a gap between the LSB and
the largest undistillable/PPT ball for the multipartite case \cite{gurvits2}.
Whether such a layering exists in the multipartite case, as opposed to the
bipartite case, remains an open problem.\newline\textbf{Acknowledgement:} The
authors would like to thank Tal Mor for useful comments. This work was
sponsored in part by the Defense Advanced Research Projects Agency (DARPA)
project MDA972-99-1-0017, and in part by the U. S. Army Research Office/DARPA
under contract/grant number DAAD 19-00-1-0172. \vspace*{-0.6cm}


\end{document}